\documentclass[twocolumn,prb,showpacs,superscriptaddress]{revtex4}

\usepackage{bbold}
\usepackage{mathptmx}
\usepackage{subfig}
\usepackage{psfrag,graphicx}
\usepackage{dcolumn}
\usepackage{amsmath,amssymb}
\usepackage{bm}
\usepackage{color}
\usepackage{latexsym}
\usepackage{epstopdf}
\usepackage{color}
\usepackage[english]{babel}
\usepackage{latexsym}
\usepackage{psfrag,graphicx}
\usepackage{amsmath}
\usepackage{amssymb}
\usepackage{amsfonts}
\usepackage{bm}
\usepackage{natbib}
\usepackage{epstopdf}
\usepackage{appendix}
\usepackage{ulem}

\definecolor{mygrey}{gray}{0.35}
\definecolor{myblue}{rgb}{0.2,0.2,0.8}
\definecolor{myzard}{cmyk}{0,0,0.05,0}
\definecolor{mywhite}{rgb}{1,1,1}
\definecolor{mywhite}{rgb}{1,1,1}
\definecolor{myred}{rgb}{1,0.,0.3}

\usepackage[colorlinks=true,citecolor=myblue,linkcolor=myred]{hyperref}

\def\ba{\begin{align}}
\def\enda{\end{align}}
\def\bi{\begin{itemize}}
\def\ei{\end{itemize}}

\def\be{\begin{equation}}
\def\ee{\end{equation}}
\def\bea{\begin{eqnarray}}
\def\eea{\end{eqnarray}}
\def\bse{\begin{subequations}}
\def\ese{\end{subequations}}

\newcommand{\ket}[1]{|{#1}\rangle}                       % ket
\newcommand{\bra}[1]{\langle {#1}|}                      % bra
\newcommand{\average}[1]{\langle {#1} \rangle}           % media < >

\newcommand{\Ignore}[1]{ }

\begin{document}
\title{Landau-Majorana-St\"uckelberg-Zener dynamics driven by coupling for two interacting qutrit systems}

\author{R. Grimaudo}
 \affiliation{ Dipartimento di Fisica e Chimica dell'Universit\`a di Palermo, Via Archirafi, 36, I-90123 Palermo, Italy}
\affiliation{ INFN, Sezione Catania, \textit{I-95123} Catania, Italy}

\author{ N. V. Vitanov }
\affiliation{Department of Physics, St. Kliment Ohridski University of Sofia, 5 James Bourchier Boulevard, 1164 Sofia, Bulgaria}

\author{A. Messina}
\affiliation{ INFN, Sezione Catania, \textit{I-95123} Catania, Italy}
\affiliation{ Dipartimento di Matematica ed Informatica dell'Universit\`a di Palermo, Via Archirafi, 34, I-90123 Palermo, Italy}

\begin{abstract}
A time-dependent two interacting spin-qutrit model is analysed and solved.
The two interacting qutrits are subjected to a longitudinal field linearly varying over time as in the Landau-Majorana-St\"uckelberg-Zener (LMSZ) scenario.
Although a transverse field is absent, we show the occurrence of LMSZ transitions assisted by the coupling between the two spin-qutrits.
Such a physical effects permits to estimate experimentally the coupling strength between the spins and allows the generation of entangled states of the two qutrits by appropriately setting the slope of the ramp.
Furthermore, the possibility of local and non-local control as well as the existence of dark states of the two qutrits have been brought to light.
Effects stemming from a noisy surrounding environment are also taken into account by introducing a random fluctuating field component as well as non-Hermitian terms in the Hamiltonian model.
\end{abstract}

\date{\today}

\maketitle

%%%%%%%%%%%%%%%%%%%%%%%%%%%%%%%%%%%%%%%%%%%%%%%%%%%%%%%%%%%%%%%%%%%%%%%%%%%
%%%%%%%%%%%%%%%%%%%%%%%%%%%%%%%%%%%%%%%%%%%%%%%%%%%%%%%%%%%%%%%%%%%%%%%%%%%
%%%%%%%%%%%%%%%%%%%%%%%%%%%%%%%%%%%%%%%%%%%%%%%%%%%%%%%%%%%%%%%%%%%%%%%%%%%
%========================================================================
%========================================================================
\section{Introduction}

Spin chains are the reference experimental scenario for quantum technology applications thanks to the possibility of entanglement generation \cite{Jurcevic,Richerme,Boness} also over long distances \cite{Sahling}.
Entanglement, indeed, is the key resource for quantum information tasks \cite{Amico} and its manipulation by field application \cite{Arnesen} is of course of fundamental importance.

In this context, a growing interest in qutrits - three-state quantum systems - should be emphasized.
Besides the obvious exponential increase of their Hilbert space, qutrits, and qudits in general, offer several advantages over qubits.
For example, among the most important applications of qutrit systems we find: optimization of the Hilbert space dimensionality vs. control complexity \cite{Greentree2004}, larger violations of nonlocality \cite{Kaszlikowski2000}, new types of quantum protocols \cite{Molina-Terriza2005} and entanglement \cite{Vaziri2002}, more secure quantum communication \cite{Cerf2002}, Bell inequalities resistant to noise \cite{Collins2002}.
Moreover, efficient protocols and methods have been developed for the manipulation of qutrits \cite{Klimov2003,Vitanov2012} and qudits \cite{Ivanov2006}.

In this respect the possibility of realizing a local application of fields on a single qudit while it interacts with other ones is of basic interest to generate physical effects in the spin chain by manipulating the single spin dynamics.
Through the Scanning Tunneling Microscopy (STM), for example, it is possible to construct atom by atom a chain of interacting nanomagnets and to manipulate the state of a single spin by applying a local magnetic field on atomic scale with a STM tip \cite{Sivkov,Yan,Wieser,Bryant,Wiebe,Tao,Lutz}.
More precisely, the field created on the single spin is an effective magnetic field stemming from the tunable exchange interaction between the target spin we wish to manipulate and the spin present on the STM tip \cite{Sivkov,Yan,Wieser,Bryant,Wiebe,Tao,Lutz}.
Such an effective field may be also time-dependent thanks to the possibility of varying the distance between the tip spin and the one in the chain \cite{Wieser}.
It is possible, for example, to create a field varying linearly in time and changing its direction \cite{Sivkov}, as in the well known Landau-Majorana-St\"uckelberg-Zener (LMSZ) scenario \cite{LMSZ}.
Thus, STM makes experimentally possible, by atomic manipulation, to control the quantum state and the quantum dynamics of a single spin while the latter is interacting with other neighbouring spins and to generate, then, delocalized effects by local field application.

The LMSZ scenario is one of the most famous and important exactly solvable time-dependent single-spin models thanks to the fact that, though its unphysical nature (infinite time duration of the physical process, implying divergence of the instantaneous energy separation as time goes on), it furnishes accurate predictions also for more realistic situations (finite times).
However, the exact solutions of the LMSZ dynamical problem exist and may be given in terms of parabolic cilinder functions \cite{Vit-Garr}.
Its popularity is confirmed also by a lot of studies, both theoretical and experimental, which have been developed aiming at generalizing the LMSZ scenario considering $N$-level systems \cite{Vasilev,SSIvanov,Pok-Sin2002,Sinitsyn1,Sinitsyn}, total crossing of bare energies \cite{Militello1} and the presence of classical and quantum noise stemming from sources of incoherences \cite{Akulin,Vitanov,Ivanov,Pok-Sin2003,Pok-Sin2004,Sin-Prok,Benza,Pok-Sun,Shilling,Kenmoe,Kenmoe1,Fai,Jipdi,Huang,Militello2,Militello3}: incoherent (mixed) states, relaxation processes (e.g., spontaneous emission) or interaction with a surrounding environment (e.g., nuclear spin bath).
Moreover, recently, the attention has been focused on double interacting spin-qubit systems subjected to LMSZ scenario \cite{VitPRL2001,Ribeiro,Ribeiro1,GVM} with the scope of identifying the signatures of the coupling in the two-spin dynamics and their potentiality for possible future applications \cite{Mhel}.

In this paper we analyse, instead, the quantum dynamics of two interacting qutrits subjected to a LMSZ ramp along the quantization axis ($\hat{z}$).
Both the manipulation of the quantum dynamics of a single high spin-value magnetic atom or molecular magnet \cite{Wieser} and the control of the interaction between qudits in a chain \cite{Sivkov,Yan} offer, indeed, the experimental background in which such systems turn out to be actual powerful building blocks for quantum information and computation tasks.

The physical interest of the work is twofold.
Firstly, we bring to light the existence of a physical effect consisting in the possibility of generating LMSZ transitions in the two-qutrit system, though a transverse constant field is absent.
This fact results possible thanks to the coupling existing between the two spin-1's which plays the role of an effective transverse field making possible avoided crossings and consequent LMSZ transitions of the two qutrit system.
Secondly, we show how such an effect may be exploited for two relevant applications: the estimation of the strength of the coupling parameters and the possibility of generating asymptotically entangled states of the two qutrits by appropriately setting the slope of the ramp.
Our symmetry-based analysis of the Hamiltonian model, usefully used in several problems \cite{GMN,GMIV,GBNM,GLSM}, allows us to consider also effects stemming from the presence of a noisy field component.

The structure of the paper is the following.
The model and the symmetry-based dynamical reduction are presented in Sec. \ref{model}.
In Sec. \ref{4D Dyn} and \ref{5D Dyn} the quantum dynamics of the two qutrits is investigated in the four- and five-dimensional dynamically invariant subspace, respectively.
In both sections we report the formal general solution of the dynamical problem and the LMSZ transition probabilities when a linearly varying ramp is applied on just one spin as well as on both the spins.
Basing on such a result, we show the possibility of local a non-local control of the dynamics of one of the two qutrits in the chain as well as physical effects related to the anisotropy of the coupling.
We bring to light moreover the existence of dark states, that is, not evolving states independently of the time-dependence of the applied fields.
Finally we discuss the modification of the LMSZ probabilities when a random fluctuating field component is present.
In Sec. \ref{Entanglement} the study of the Negativity as measure of Entanglement between the two qutrits is developed and the possibility of generating entangled states of the two spin-1's through a LMSZ process is analysed.
Finally, conclusive remarks and perspectives may be found in the last section \ref{Conclusions}.

\section{The Model}\label{model}

Let us consider the following model of two interacting qutrits subjected to local time-dependent fields
\begin{eqnarray} \label{Hamiltonian}
H&=&
\hbar\omega_{1}\hat{\Sigma}_{1}^{z}+\hbar\omega_{2}\hat{\Sigma}_{2}^{z}
+\gamma_{x}\hat{\Sigma}_{1}^{x}\hat{\Sigma}_{2}^{x}
+\gamma_{y}\hat{\Sigma}_{1}^{y}\hat{\Sigma}_{2}^{y}
+\gamma_{z}\hat{\Sigma}_{1}^{z}\hat{\Sigma}_{2}^{z}
\end{eqnarray}
where $\omega_i$ ($i=1,2$) are the characteristic frequencies of the two qutrits and $\gamma$s are the different energy contributions stemming from the coupling between the two three-level systems.
The Pauli operators $\hat{\Sigma}_i^k$ ($k=x,y,z$) for a spin-1 system are related with the spin-1 operator components as 
\begin{equation}\label{Relations Pauli operators-Angular momentum spin 1}
\hat{S}_i^x = {\hbar \over \sqrt{2}} \hat{\Sigma}_i^x, \quad \hat{S}_i^y = {\hbar \over \sqrt{2}} \hat{\Sigma}_i^y,
 \quad \hat{S}_i^z = \hbar \hat{\Sigma}_i^z.
\end{equation}
Our scope is to study a Landau-Majorana-St\"uckelberg-Zener (LMSZ) scenario for the two qutrits and analyse how the coupling between them and a noisy component of the magnetic field affect their dynamics.

In Ref. \cite{GMIV} it was shown that two dynamically invariant Hilbert subspaces exist: one of dimension four spanned by $\{\ket{10},~\ket{01},~\ket{0-1},~\ket{-10}\}$ and the other one of dimension five spanned by $\{\ket{11},~\ket{1-1},~\ket{00},~\ket{-11},~\ket{-1-1}\}$.
They are related to the two eigenvalues ($\pm 1$) of the constant of motion
\begin{equation}\label{Cos Form of K}
\hat{K} = \cos(\pi\hat{\Sigma}_{\rm tot}^z),
\end{equation}
where $\hat{\Sigma}_{\rm tot}^z = \hat{\Sigma}_{1}^{z} + \hat{\Sigma}_{2}^{z}$ is the total spin of the composed system along the $z$ direction.
It is worth to emphasize, at this point, that the Hamiltonian model keeps its symmetry also for two larger spin systems, that is, for two interacting spins $\mathbf{\hat{J}}_1$ and $\mathbf{\hat{J}}_2$.
In such a case, it is always possible to decompose the dynamical problem into two sub-problems related to the two dynamically invariant subspaces linked to the two eigenvalues ($1$ and $-1$) of the constant of motion $\cos[\pi(\hat{J}_{1}^z+\hat{J}_{2}^z)]$.
However, for larger spin systems, the sub-dynamics could be very difficult to solve due to the high degeneracy of both eigenvalues.

In this respect, in \cite{GMIV} an important property of the two-qutrit system was discovered, which is of basic importance for our analysis and to get exact solutions of the dynamical problem.
The Hamiltonian governing the two-qutrit dynamics in the four-dimensional subspace may be written in terms of two \textit{non interacting} qubits as follows
\begin{equation}\label{4x4 block as two spin-1/2}
H_{-}=H_{1} \otimes \hat{\mathbb{1}}_{2}+\hat{\mathbb{1}}_{1} \otimes H_{2},
\end{equation}
with
\begin{equation}\label{H1 and H2}
H_{1}=\frac{\hbar\Omega_+}{2}\hat{\sigma}_{1}^{z}+
\gamma_-\hat{\sigma}_{1}^{x},
%+(\gamma_{xy}+\gamma_{yx})\hat{\sigma}_{1}^{y},
\qquad
H_{2}=\frac{\hbar\Omega_-}{2}\hat{\sigma}_{2}^{z}+
\gamma_+\hat{\sigma}_{2}^{x}
%-(\gamma_{xy}-\gamma_{yx})\hat{\sigma}_{2}^{y},
\end{equation}
where $\hat{\sigma}^k$ ($k=x,y,z$) are the standard Pauli matrices and we set $\Omega_\pm=\omega_1\pm\omega_2$ and  $\gamma_\pm=\gamma_x\pm\gamma_y$.
The mapping at the basis of such a rewriting is
\begin{equation}\label{Mapping}
\begin{aligned}
\ket{10} & \hspace{0,25cm} \leftrightarrow \hspace{0,25cm} \ket{++},\\
\ket{01} & \hspace{0,25cm} \leftrightarrow \hspace{0,25cm} \ket{+-},\\
\ket{0-1} & \hspace{0,25cm} \leftrightarrow \hspace{0,25cm} \ket{-+},\\
\ket{-10} & \hspace{0,25cm} \leftrightarrow \hspace{0,25cm} \ket{--}.
\end{aligned}
\end{equation}

The Hamiltonian governing the five dimensional subspace, instead, under the following conditions
\begin{equation}\label{Specific parameters conditions}
\begin{aligned}
\gamma_z=0, \qquad
\gamma_x = \gamma_y = \gamma/2,
\end{aligned}
\end{equation}
is reduced to the following block-diagonal form
\begin{equation}\label{H+}
H_+=
\left(
\begin{array}{ccccc}
 \hbar\text{$\Omega_+$} & 0 & 0 & 0 & 0 \\
 0 & \hbar\text{$\Omega_-$} & \gamma  & 0 & 0 \\
 0 & \gamma  & 0 & \gamma & 0 \\
 0 & 0 & \gamma  & -\hbar\text{$\Omega_- $} & 0 \\
 0 & 0 & 0 & 0 & -\hbar\text{$\Omega_+$}
\end{array}
\right).
\end{equation}
The three-dimensional middle block possesses an su(2) structure and hence can be written in terms of spin variables of a fictitious spin-1, namely
\begin{equation}\label{H3}
H_3=
\gamma \hat{\Sigma}^{x}+\hbar\Omega_-\hat{\Sigma}^{z}.
\end{equation}
We emphasize that the choice $\gamma_z=0$ is necessary to get an su(2)-symmetry structure of the matrix within the three-dimensional subspace.
This choice, however, does not alter the four-dimensional sub-dynamics since $H_1$ and $H_2$ in Eq. \eqref{H1 and H2} do not depend on $\gamma_z$.

Now, we want to study the two interacting qutrits when they are subjected to time-dependent fields, $\omega_1(t)$ and $\omega_2(t)$.
To this end we stress that the results and the analysis reported before in Ref. \cite{GMIV} are still valid also when we consider time-dependent fields and, more generally, when all the Hamiltonian parameters depend on time.
This is due to the fact that the Hamiltonian structurally commutes with the constants of motion independently of its time-dependence.
In the following we show that we are able to construct formally the time evolution operator for both four- and five-state subdynamics.
In particular, we analyse the case in which the $z$-magnetic field is a ramp as in the LMSZ scenario.
We are interested in revealing intriguing dynamical effects stemming from the homogeneity or heterogeneity of both the coupling parameters and the two fields.
In addition, we want to exploit our symmetry-based approach to take into account the influence of a surrounding environment by considering a random fluctuating field component.

\section{Four-Dimensional Subdynamics} \label{4D Dyn}

\subsection{General Solution}

We may formally write the time evolution operator $U_j$ ($j=1,2$) related to $H_j$, solution of the Schr\"odinger equation $i\hbar\dot{U}_j=H_jU_j$, as follows
\begin{equation}
U_j=
\begin{pmatrix}
a_j & b_j \\
-b_j^* & a_j^*
\end{pmatrix},
\end{equation}
where $a_j$ and $b_j$ are time-dependent Cayley-Klein parameters satisfying $|a_j|^2+|b_j|^2=1$.
The time evolution operator $U_-$, satisfying the Schr\"odinger equation $i\hbar\dot{U}_-=H_-U_-$, then reads
\begin{equation}
\begin{aligned}
U_-=U_1 \otimes U_2=
\left(
\begin{array}{cccc}
 a_1 a_2 & a_1 b_2 & b_1a_2  & b_1 b_2 \\
 -a_1 b_2^* & a_1 a_2^* & -b_1 b_2^* & b_1 a_2^* \\
 -b_1^*a_2  & -b_1^*b_2  & a_1^*a_2  & a_1^*b_2  \\
 b_1^* b_2^* & -b_1^*a_2^*  & -a_1^* b_2^* & a_1^* a_2^* \\
\end{array}
\right).
\end{aligned}
\end{equation}
The mathematical expressions of $a_j(t)$ and $b_j(t)$ depend on the time-dependence of the two local magnetic fields $\omega_1(t)$ and $\omega_2(t)$.

\subsection{STM Scenario}

\subsubsection{Local Dynamics}

\begin{figure}[htp]
\begin{center}
\includegraphics[width=0.3\textwidth]{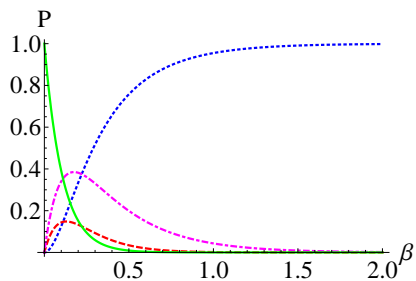}
\captionsetup{justification=raggedright,format=plain,skip=4pt}%
\caption{(Color online) a) Asymptotic LMSZ probabilities [Eq. \eqref{MLSZ Trans Prob 0-1 to 01}] of finding the two qutrits in the state $\ket{10}$ (blue dotted line), $\ket{01}$ (magenta dot-dashed line), $\ket{0-1}$ (red dashed line) and $\ket{-10}$ (green full line), when they start from the state $\ket{-10}$ for $\gamma_x\neq\gamma_y$, $\beta=\beta_+$ and $\beta_+/\beta_-=2$.}\label{fig:PLZ4D}
\end{center}
\end{figure}

We firstly analyse the case of a single local $z$-magnetic field $B_z(t)$ applied on the first spin consisting in a LMSZ ramp, such that
\begin{equation}
\hbar\omega_1(t)={\alpha} t, \quad t\in (-\infty,\infty),
\end{equation}
where $\alpha$ is considered a positive real number and rules the adiabaticity of the process since $\dot{B}_z\propto\alpha$.
Let us consider the case of an excitation present in the system and localized in one of the two qutrits, say the second spin; in this case the initial state of the two qutrits (fictitious qubits) is $\ket{-10}$ ($\ket{--}$).
In this instance, each fictitious spin-1/2 is subjected to a LMSZ scenario with $\omega_1(t)$ as longitudinal magnetic field and a constant (effective) transverse magnetic field determined by the coupling parameters [see Eq. \eqref{H1 and H2}].
In this way, the first and second fictitious spin-1/2 have the probability to make the transition to the up-state, respectively
\begin{equation}\label{P1}
P_1=1-\exp\{-2\pi\beta_-\},
\end{equation}
and
\begin{equation}\label{P2}
P_2=1-\exp\{-2\pi\beta_+\},
\end{equation}
with $\beta_\pm=\gamma_\pm^2/\hbar\alpha$.
Thus, the joint probability for the two fictitious spin-1/2's to be found in the state $\ket{++}$, $\ket{+-}$ and $\ket{-+}$, starting from $\ket{--}$, are respectively
\begin{equation}\label{MLSZ Trans Prob 0-1 to 01}
\begin{aligned}
P_1 P_2, \qquad P_1 (1-P_2), \qquad (1-P_1) P_2,
\end{aligned}
\end{equation}
being nothing but the probability of finding the two qutrits in the state $\ket{10}$, $\ket{01}$ and $\ket{0-1}$, respectively.
We know that in the standard LMSZ scenario applied on a single spin-qubit, the transverse field couples the two levels and is then responsible of the avoided crossing.
It is worth noticing that, in our case, the transverse field role is played by the coupling existing between the two qutrits, as it is clear by the two Hamiltonians in Eq. \eqref{H1 and H2}.
Hence, we may reproduce adiabatic conditions by appropriately setting the ratio between the longitudinal fields and the coupling parameters in order to have a full LMSZ transition of the two fictitious spin-1/2's.
The three probabilities in Eq. \eqref{MLSZ Trans Prob 0-1 to 01} are reported in Fig. \ref{fig:PLZ4D} against the parameter $\beta=\beta_+$ for $\beta_+/\beta_-=2$.
In this case we are realizing a local control of the dynamics of the first qutrit, leaving the other one unaltered.
For a complete LMSZ transition, indeed, the first qutrit accomplishes the LMSZ transition $\ket{-1} \rightarrow \ket{1}$, while the second qutrit's state does not change.

Analogously, we may consider the excitation initially localized in the first spin-1, so that the two qutrits start from the state $\ket{0-1}$.
In this instance the two-qutrit system is asymptotically driven to the state $\ket{01}$ and the probability of the related transition acquires the same expression as the previous one in Eq. \eqref{MLSZ Trans Prob 0-1 to 01}.
It is worth noticing that in this case we generate a LMSZ transition from $\ket{-1}$ to $\ket{1}$ in the second spin, by applying a local magnetic field only on the first qutrit which, instead, remains in its initial state.
Such a circumstance, thus, may be identified as the achievement of a non-local control of the second qutrit.

\subsubsection{State Transfer between the Qutrits}

Another interesting effect to be highlighted is the possibility of realizing a state transfer between the two qutrits.
Indeed, if the two qutrits (fictitious qubits) are initialized in the state $\ket{-10}$ ($\ket{--}$) and we assume $\gamma_x=\gamma_y$, the transition probability of the first fictitious spin-1/2 is forbidden, while the second one passes to $\ket{+}$ with probability $P=P_2$.
In this way, the two qutrits (fictitious qubits) reach the state $\ket{0-1}$ ($\ket{-+}$) having interchanged their initial state.
The same effect is present if the two qutrits are initially prepared in $\ket{10}$ passing to $\ket{01}$.
In such a case, the transitions between the states of the two qutrit system in the four dimensional subspace are different since the condition $\gamma_x=\gamma_y$ introduces a further symmetry in the model related to the commutation of $H$ with $\hat{\Sigma}_{\text{tot}}^z$.
This fact generates, in the subspace under scrutiny, the existence of other two dynamically invariant subspaces related to the eigenvalues of $\hat{\Sigma}_{\text{tot}}^z$.
It is easy to verify that, this time the two qutrits starting from $\ket{-10} $ ($\ket{10}$) can be asymptotically found only in the state $\ket{0-1}$ ($\ket{01}$).

At the light of the STM scenario, the physical effects previously discussed and analytically derived are of relevant interest.
They show, indeed, that the presence of the coupling between the two qutrits allows us to manipulate the dynamics of the whole two-qutrits chain by the application of a single local magnetic field on one of the two spins, being exactly one of the task of the application of the STM technique.
Moreover, the previous examples brought to light that, by studying the kind of transitions occurring in the two-qutrit system, we may get information about the coupling parameters determining the symmetries of the Hamiltonian.

\subsubsection{Effects of Environment}

We wish to show now that the mapping of the two-qutrit dynamics into that of two decoupled spin-1/2's in the four-dimensional subspace is useful not only to solve exactly the problem in ideal conditions, but also to take into account possible external influences due to the action of a surrounding environment, such as nuclear spin bath.
In Ref. \cite{Huang}, for example, it is experimentally demonstrated that decoherence effects in the dynamics of a NV center in diamond (consisting in a three-level system), subjected to a LSZ interferometer, comes from the dipolar interaction of the system with the surrounding $^{13}$C nuclear spins random fluctuating at room temperature.
Such external influences may be theoretically regarded, for example, as noise in the magnetic field component.
In Ref. \cite{Pok-Sin2003} the authors study the dynamics of a spin $S$ subjected to a noisy LMSZ scenario.
The noisy time-dependent magnetic field $\eta(t)$ is considered only in the $z$ direction and characterized by a time correlation function of the form $\average{\eta(t)\eta(t')}=2\Gamma\delta(t-t')$.
Reference \cite{Huang} experimentally legitimates such an assumption; in that case, indeed, the authors shows as the transverse fluctuations can be neglected.
In such a way the noisy component cannot generates transitions between the different states but it leads only to loss of coherence.
In Ref. \cite{Pok-Sin2003}, the authors show how the LMSZ transition probability is affected by the presence of such a noisy magnetic field in the case of a spin-1/2, a spin-1 and a spin-3/2.
For a spin-1/2 and for large values of $\Gamma$ we have asymptotically
\begin{equation}
P_-^+={1-\exp\{-2\pi g^2/\hbar\alpha\} \over 2},
\end{equation}
where $g$ is the energy contribution due to the coupling of the spin-1/2 with the constant transverse magnetic field.
We see that the transition probability does not depend on the specific value of $\Gamma$, provided that $\Gamma$ is large.
Moreover, it is important to note that the effect of the noise is to hinder the transition.
Indeed, in the most convenient case, that is for $g^2/\hbar\alpha\gg 1$, the system reaches at most an equally populated condition of the two states. 
This is of particular interest for us since we have shown that the transition of the two qutrits studied before can be reduced to the LMSZ transition of a spin-1/2.
Then, it means that the result previously reported can be exploited in our case to find the corrected LMSZ transition probability for the two qutrits when the field is affected by a noisy component.
For example, if $\gamma_x \neq \gamma_y$, the probability in Eq. \eqref{MLSZ Trans Prob 0-1 to 01} becomes $P_{12}/4$, reasonably meaning that, under the effect of noise, we reach an equally populated condition of the four states involved in the subdynamics under scrutiny.
Analogously, if $\gamma_x = \gamma_y$, had the two qutrits started form $\ket{-10}$ we get the probability $P_2/2$ of transition to the state $\ket{0-1}$, reaching this time an equally populated condition between these two states.

Such observation is based on the fact that, adding the noisy component $\eta(t)$ to the field applied to the first qutrit, nothing changes in the dynamics-decoupling procedure.
The Hamiltonian transformation is completely unaffected since the only difference consists in a redefinition of the longitudinal field.
In this way, what we obtain is an effective $z$-field for the two fictitious spin-1/2's supplemented by a random field component.
Thus, also in this case, we may reduce the two-qutrit dynamical problem into the analysis of the quantum dynamics of two decoupled spin-1/2's.

In this respect, it is worth pointing out that the argument previously exposed continues to be valid also when we consider the possibility that the exited states $\ket{0}$ and $\ket{1}$ of the two qutrits decay irreversibly out of the system by some mechanism.
Let us suppose that the spontaneous emission from the exited states to the ground one is negligible and that the two decay rates for the state $\ket{0}$ and $\ket{1}$ are $\tilde{\Gamma}$ ($\tilde{\Gamma}'$) and $2\tilde{\Gamma}$ ($2\tilde{\Gamma}'$), respectively, for the first (second) qutrit.
It is easy to see that the analysis of such a scenario is equivalent, up to add a constant imaginary term, to phenomenologically introduce the non-Hermitian terms $i\tilde{\Gamma}\hat{\Sigma}_1^z$ and $i\tilde{\Gamma}'\hat{\Sigma}_2^z$ in our Hamiltonian model.
Also this time we have a simple redefinition of the parameters in front of the operators $\hat{\Sigma}_1^z$ and $\hat{\Sigma}_2^z$ without altering the symmetries possessed by the Hamiltonian $H$.
Therefore, in such a case, within the four-dimensional subspace the two-qutrit dynamics may be described in terms of two decoupled two-level systems subjected to effective external fields and characterized by decaying states.
Several results have been reported for a single qubit with a decaying state subjected to the LMSZ scenario \cite{Akulin,Vitanov,Ivanov}.
Precisely, it has been proved that, on the one hand, in the standard (ideal) LMSZ scenario, the decay rate influences only its the time-history of the transition probality but not its asymptotic value \cite{Akulin}; on the other hand, in the more realistic LMSZ scenario characterized by a limited time-window, the exited state population exhibits a dependence on the decay rate \cite{Vitanov}.
We emphasize that even such results allow to make quantitative predictions on the LMSZ transition probabilities for the system under scrutiny.

\subsection{Local Fields}\label{Local Fields}

Now, we want to discuss the possibility of applying local fields on both the qutrits.
Let us consider, firstly, the case
\begin{equation}
\omega_1(t)=\omega_2(t)={\alpha} t/2,
\end{equation}
with $t$ going from $-\infty$ to $+\infty$.

In this case, the Hamiltonians of the two fictitious spin-1/2's, through which we describe effectively the dynamics of the two qutrits in the four dimensional subspace, read
\begin{equation}
H_{1}=\hbar\Omega_+(t)\hat{\sigma}_{1}^{z}+
\gamma_-\hat{\sigma}_{1}^{x},\qquad
H_{2}=\gamma_+\hat{\sigma}_{2}^{x},
\end{equation}
with $\Omega_+(t)=\alpha t$.
We see that the second fictitious spin-1/2 is subjected only to a magnetic field in the $x$-direction, while the first one is subjected to standard Landau-Zener scenario.
As before, the role of the external transverse constant field is effectively played by the coupling existing between the two spins.

\subsubsection{Determination of $\gamma$s}

We study now the instance in which only one excitation is present in the system, equally shared by the two qutrits.
We consider, then, the entangled state $(\ket{-10}+\ket{0-1})/\sqrt{2}$ as initial condition.
By the mapping in Eq. \eqref{Mapping}, such a state, rewritten in terms of the two spin-1/2 states, acquires the form
\begin{equation}
\ket{-} \otimes {\ket{+}+\ket{-} \over \sqrt{2}}.
\end{equation}

It is easy to see that the second spin does not change its state in time since the latter is an eigenvalue of $H_2$.
The first spin, instead, evolves according to the LMSZ dynamics, so that the probability to find it in the opposite state $\ket{+}$ at very large time instants ($t \rightarrow \infty$) is $P_1$.
%\begin{equation}\label{LZ Trans Prob}
%P_-^+=1-\exp\{-2\pi\beta\},
%\end{equation}
%with $\beta=\gamma_-^2/\hbar\alpha$.
Of course, it expresses too the probability of the two spin-1/2's to be found in the state $\ket{+} \otimes {\ket{+}+\ket{-} \over \sqrt{2}}$.
The relevant point is that, in view of Eq. \eqref{Mapping}, it provides the probability for the two qutrits of reaching the state
\begin{equation}
{\ket{10}+\ket{01} \over \sqrt{2}}.
\end{equation}
Thus, if $\beta_-\gg 1$, through the linear ramp we have created an excitation in the system.
It is important to underline that such a transition depends strongly on the coupling parameters between the two qutrits, since their difference constitute the effective transverse magnetic field entering in the expression of the LMSZ parameter $\beta_-$.
Indeed, if the two parameters are equal or very close, the transition is forbidden, while, if they are opposite, the transition probability reaches its maximum efficiency.
This suggests us that, choosing at will $\alpha$ and studying the characteristic time of the transition, we may get information about the value of $\gamma_-$.

If we now consider
\begin{equation}
\omega_1(t)=-\omega_2(t)=\alpha t/2
\end{equation}
and the two qutrits initially prepared in the state $({\ket{-10}+\ket{0-1}) / \sqrt{2}}$, we get a specular dynamics.
That is, the first fictitious spin-1/2, subjected only to a static $x$-magnetic field ($H_1=\gamma_-\hat{\sigma}_1^x$), does not evolve, while the second fictitious spin-1/2 makes a transition from $\ket{-}$ to $\ket{+}$ (being $H_2=\hbar\alpha t\hat{\sigma}_{2}^{z}+\gamma_+\hat{\sigma}_{2}^{x}$).
Studying such a transition, this time, we get information about $\gamma_+$ since it rules the characteristic time of such a transition.
Finally, by comparing the two values of $\gamma_+$ and $\gamma_-$ we may estimate the original coupling parameters of the two qutrits $\gamma_x$ and $\gamma_y$.

\subsubsection{Dark States}

We emphasize that, under the conditions $\gamma_x=\gamma_y=\gamma/2$ and $\omega_1(t)=\omega_2(t)=\omega(t)/2$ (unique homogeneous magnetic field), the following four states
\begin{equation}
\ket{\psi_{1/2}^0}={\ket{10}\pm\ket{01} \over \sqrt{2}}, \qquad \ket{\psi_{3/4}^0}={\ket{-10}\pm\ket{0-1} \over \sqrt{2}}
\end{equation} 
result steady states independently of the time dependences of the magnetic field.
This may be easily understood in terms of the two spin-1/2's.
Indeed, the second spin-1/2 is in an eigenstate [${(\ket{+}\pm\ket{-}) / \sqrt{2}}$] of its constant Hamiltonian $H_2=\gamma\hat{\sigma}^x$ and evolves trivially, only acquiring the phase factor $\exp\{-i\gamma t/\hbar\}$; the first fictitious spin-1/2, instead, (being in the state $\ket{\pm}$) keeps only the phase factor $\exp\{-i\int_0^t\omega(t)dt\}$ since its Hamiltonian $H_1=\hbar\omega(t)\hat{\sigma}_1^z$ does not mix the two standard basis states.
This means that for these four states we have ($j=1 \dots 4$)
\begin{equation}
\begin{aligned}
& H(t)\ket{\psi_j^0}=E_j(t)\ket{\psi_j^0}, \\
& E_{1/2}(t)=\omega(t)\pm\gamma, \quad E_{3/4}(t)=-E_{2/1}(t) 
\end{aligned}
\end{equation}
implying
\begin{equation}
\ket{\psi_j(t)}=\exp\left\{-i\int_0^t dt'E(t')/\hbar\right\} \ket{\psi_j^0}.
\end{equation}
It is easy to see that, considering the time-independent case, such states result to be the eigenstates of the Hamiltonian \cite{GMIV}.
So, this model, in this specific case, presents a peculiar characteristic consisting in maintaining its steady states also when the Hamiltonian parameters are time-dependent.
A remarkable consequence of this circumstance is that the following class of states $\rho_0=\sum_jp_j\ket{\psi_j^0}\bra{\psi_j^0}$ ($\sum_jp_j=1$), comprising e.g. the thermal state ($p_j=\exp\{-E_j/k_BT\}$, $k_B$ and $T$ being the Boltzman constant and the Temperature, respectively), do not evolve in time, that is
\begin{equation}\label{Stationary Mixed States}
\rho(t)=\sum_jp_j\ket{\psi_j(t)}\bra{\psi_j(t)}=\sum_jp_j\ket{\psi_j^0}\bra{\psi_j^0}=\rho_0.
\end{equation}
Therefore, any physical observable calculated for such class of states exhibit a constant value in time.
We can call such states `dark states' since, under the conditions written before, they are unaffected by both the coupling and the longitudinal time-dependent field, also when the latter presents a random fluctuating behaviour.

Analogously, if we have $\gamma_x=-\gamma_y$ and $\omega_1(t)=-\omega_2(t)$ the four dark states are
\begin{equation}
{\ket{10}\pm\ket{0-1} \over \sqrt{2}}, \qquad {\ket{01}\pm\ket{-10} \over \sqrt{2}}.
\end{equation}
Finally, we emphasize that the previous results are not restricted to the LMSZ scenario, but they are valid whatever the time-dependence of the field is.

\section{Five-Dimensional Subdynamics} \label{5D Dyn}

\subsection{General Solution}

In the second section we saw that the central block of $H_+$ in Eq. \eqref{H+} has an su(2) structure and then it is interpretable as the Hamiltonian of a (fictitious) spin-1 subjected to (fictitious as well) magnetic fields (see Eq. \eqref{H3}).
It is well known that the time evolution operator related to a 3x3 su(2) Hamiltonian may be put in the following form \cite{Hioe}
\begin{equation}\label{U3}
U_3=
\begin{pmatrix}
a_3^{2} & \sqrt{2}a_3b_3 & b_3^{2} \\
-\sqrt{2}a_3b_3^{*} & |a_3|^2-|b_3|^2 & \sqrt{2}a_3^{*}b_3 \\
{b_3^{*}}^{2} & -\sqrt{2}a_3^{*}b_3^{*} & {a_3^{*}}^{2}
\end{pmatrix},
\end{equation}
where $a_3$ and $b_3$ are two time-dependent parameters, solution of the analogous dynamical problem for a single spin-1/2.
In other words, $a_3$ and $b_3$ may be found by solving the dynamical problem of a single spin-1/2 subjected to the same magnetic field acting upon the fictitious spin-1.
Thus, we may formally write the time evolution operator $U_+$, solution of the Schr\"odinger equation $i\hbar\dot{U}_+=H_+U_+$, as follows
\begin{equation}\label{U+}
\begin{aligned}
&U_+=\\
&\left(
\begin{array}{ccccc}
 e^{-{i\over\hbar}\int\Omega_+} & 0 & 0 & 0 & 0 \\
 0 & a_3^{2} & \sqrt{2}a_3b_3 & b_3^{2} & 0 \\
 0 & -\sqrt{2}a_3b_3^{*} & |a_3|^2-|b_3|^2 & \sqrt{2}a_3^{*}b_3 & 0 \\
 0 & {b_3^{*}}^{2} & -\sqrt{2}a_3^{*}b_3^{*} & {a_3^{*}}^{2} & 0 \\
 0 & 0 & 0 & 0 & e^{{i\over\hbar}\int\Omega_+}
\end{array}
\right).
\end{aligned}
\end{equation}

\subsection{Dark States}

First of all, it is important to underline that also for the five-dimensional subdynamics we have dark states.
Indeed, if the two qutrits are initially prepared in $\ket{11}$ or $\ket{-1-1}$, independently of the time-dependence of the $z$-magnetic field, the two-qutrit system remains in its initial state, also if the magnetic field component randomly fluctuates remaining along the $z$-direction.
Moreover, if we consider the case $\omega_1(t)=\omega_2(t)$, also a generic state belonging to the three-dimensional subspace, namely
\begin{equation}\label{Gen State 3dim Space}
c_1\ket{1-1}+c_2\ket{00}+c_3\ket{-11},
\end{equation}
is completely unaffected by the presence of time-dependent magnetic fields, since in this instance $\Omega_-(t)=0$ and the Hamiltonian governing the three-dymensional dynamics is simply $H_3=\gamma\hat{\sigma}^x$.
Such states, then, evolves only under the action of the coupling between the two qutrits.
It means then that the three eigenstates of $\hat{\Sigma}^x$ rewritten in terms of two-qutrit states
\begin{equation}\label{Steady Mixed States}
\begin{aligned}
\ket{\psi_5^0}&={\ket{1-1}+\sqrt{2}\ket{00}+\ket{-11} \over 2},
\\
\ket{\psi_6^0}&={\ket{1-1}-\ket{-11} \over \sqrt{2}}
\\
\ket{\psi_7^0}&={\ket{1-1}-\sqrt{2}\ket{00}+\ket{-11} \over 2}
\end{aligned}
\end{equation}
result steady state of the two-qutrit system also when a unique homogeneous time-dependent field is applied on the two spin-1's.
Consequently, every classical mixture of these three states does not evolve and every physical quantity related to this state is constant in time.
Given that the states in Eq. \eqref{Stationary Mixed States} have the same property under the same conditions ($\omega_1(t)=\omega_2(t)$ and $\gamma_x=\gamma_y$), we may conclude that, in this scenario, the thermal state of the system and, more in general, every mixture involving the steady states $\ket{11}$, $\ket{-1-1}$ and the ones in Eqs. \eqref{Stationary Mixed States} and \eqref{Steady Mixed States}, namely
\begin{equation}
\rho=k_1\ket{11}\bra{11}+\sum_{j=1}^7p_j\ket{\psi_j^0}\bra{\psi_j^0}+k_2\ket{-1-1}\bra{-1-1},
\end{equation}
such that $k_1+k_2+\sum_jp_j=1$, is a stationary state of the two-qutrit system.

\subsection{STM Scenario and LMSZ Transition Probabilities}

\begin{figure}[htp]
\begin{center}
{\includegraphics[width=0.3\textwidth]{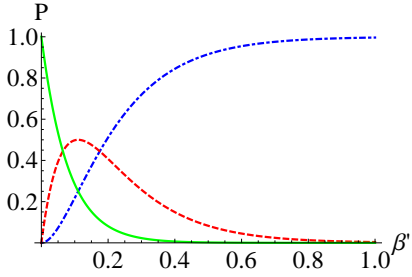}}
\captionsetup{justification=raggedright,format=plain,skip=4pt}%
\caption{(Color online) Asymptotic LMSZ probabilities [Eq. \eqref{LMSZ Tr Pr 3-lev}] of finding the two qutrits in the state $\ket{1-1}$ (blue dot-dashed line), $\ket{00}$ (red dashed line) and $\ket{-11}$ (green full line), when they start from the state $\ket{-11}$ for $\gamma_x=\gamma_y$.}\label{fig:PLZ3D}
\end{center}
\end{figure}

We investigate now the STM experimental scenario characterized by a single local magnetic field on the first spin-1, namely $\omega_1(t)=\alpha t$, and the two qutrits initialized in the state $\ket{1-1}$.
In this case the two-qutrit system behaves effectively like a three-level system (spin-1) subjected to a LMSZ ramp with an effective constant transverse magnetic field related to the coupling constant $\gamma$.
For such a time-dependent scenario, the transition probabilities, from $\ket{1-1}$ to the other two states $\ket{00}$ and $\ket{-11}$, may be found analytically.
Indeed, at the light of the spin-1 - spin-1/2 transition probability relationship based on the SU(2) group structure, for large time instants, we have
\begin{equation}\label{LMSZ Tr Pr 3-lev}
\begin{aligned}
P_{-1}^{+1}=P_3^2, \quad
P_{-1}^{0}=2P_3(1-P_3), \quad
P_{-1}^{-1}=(1-P_3)^2,
\end{aligned}
\end{equation}
where $P_3=(1-e^{-2\pi\beta'})$ and $\beta'=2\gamma^2/\hbar\alpha$.
Also in this case, we appreciate how the coupling between the two qutrits is responsible of an avoided crossing and a consequent full adiabatic LMSZ transition for the fictitious spin-1.
In the previous expressions we have labelled with -1, 0 and 1 the states $\ket{1-1}$, $\ket{00}$ and $\ket{-11}$, respectively.
The plots of the asymptotic probabilities are reported in Fig. \ref{fig:PLZ3D} against the coupling-dependent LMSZ parameter $\beta'$.
We see that the interplay between the coupling parameter $\gamma$ and the ramp of the magnetic field $\alpha$, defining $\beta'$, deeply influences the transition probability.
For high values of the parameter $\beta'$ we get a complete LMSZ transition of both the spins, getting, also this time, a state transfer between the two qutrits.
This means that, measuring the state of the system and varying the ramp $\alpha$, we may estimate the parameter $\gamma$ determining the strength of coupling between the two qutrits.

\subsection{Noise Effects}

We consider now the field along the $z$ axis affected by the random fluctuating contribution we saw in the previous section.
We may exploit again the results reported in Ref. \cite{Pok-Sin2003} where the authors solved the dynamical problem of a noisy ramp in a LMSZ scenario also for a spin-1.
In such a case, the transition probabilities affected by a noisy field component along the $z$-axis and characterized by the following time-correlation function $\average{\eta(t)\eta(t')}=2\Gamma\delta(t-t')$, become
\begin{equation}
\begin{aligned}
P_{-1}^{+1}&={1\over 6}(2+e^{-3\pi\beta'}-3e^{-\pi\beta'}), \\
P_{-1}^{0}&={1\over 3}(1-e^{-3\pi\beta'}), \\
P_{-1}^{-1}&={1\over 6}(2+e^{-3\pi\beta'}+3e^{-\pi\beta'}).
\end{aligned}
\end{equation}
Also these expressions, valid for large values of $\Gamma$, are independent of the value of the same $\Gamma$.
We see that, also this time, the main effect of the noise is to hinder the transition generating at most equally populated states when $\beta' \gg 1$.
In this way, we brought to light how the symmetry-based analysis of the model reported in the second sections plays a key role for disclosing the exact quantum dynamics of the two interacting qutrits subjected to time-dependent magnetic fields, both in ideal and more realistic conditions.

\section{Entanglement} \label{Entanglement}

The negativity, introduced by G. Vidal and R. F. Werner in \cite{Vid-Wer}, of a two-qutrit system described by the density matrix $\rho$ reads \cite{Rai-Luthra}
\begin{equation}\label{General Negativity}
\mathcal{N}_\rho = {||\rho^{T_B} ||_1 - 1 \over 2},
\end{equation}
where $\rho^{T_B}$ is the partial transpose of the matrix $\rho$ with respect to the subsystem $B$.
The symbol $|| \cdot ||_1$ is the trace norm which, for a hermitian matrix, results in the sum of the absolute values of the negative eigenvalues of ${\rho}^{T_B}$ which is hermitian and such that $\text{Tr}\{ \rho^{T_B} \}=1$.
The range of values of $\mathcal{N}_{\rho}$ is $[0,1]$ \cite{Rai-Luthra} and its calculation is independent of the factorized orthonormal basis in which the matrix $\rho$ is represented as well as of the subsystem with respect to which we calculate the partial transpose, since $(\rho^{T_A})^T =\rho^{T_B}$ and $||X||_1=||X^T||_1$ for any operator $X$.

\subsection{Four-Dimensional Sub-Dynamics}

For our two-qutrit system, it has been proved \cite{GMIV} that the negativity for a generic pure as well as mixed state belonging to the four dimensional subspace possesses the upper bound $\mathcal{N}=1/2$.
In case of a generic pure state $\ket{\Psi}=w_1\ket{10}+w_2\ket{01}+w_3\ket{0-1}+w_4\ket{-10}$, the Negativity acquires indeed the simple form \cite{GMIV}
\begin{equation}\label{Neg 4x4}
\mathcal{N}=\sqrt{x(1-x)}, \qquad x=|w_1|^2+|w_4|^2.
\end{equation}

\begin{figure}[htp]
\begin{center}
\subfloat[][]{\includegraphics[width=0.22\textwidth]{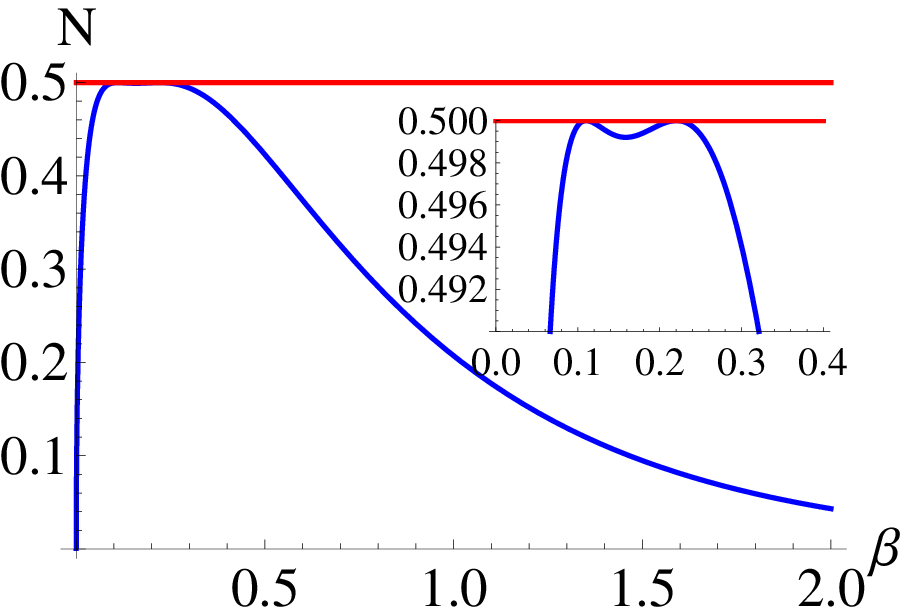}\label{fig:PbetaCurve4}}
\qquad
\subfloat[][]{\includegraphics[width=0.22\textwidth]{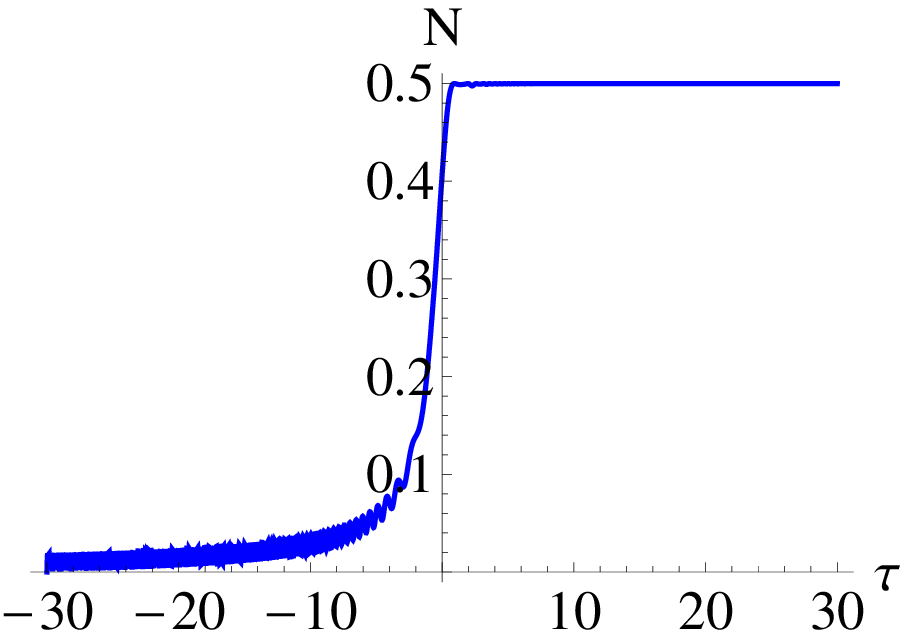}\label{fig:N4beta011}}
\qquad
\subfloat[][]{\includegraphics[width=0.22\textwidth]{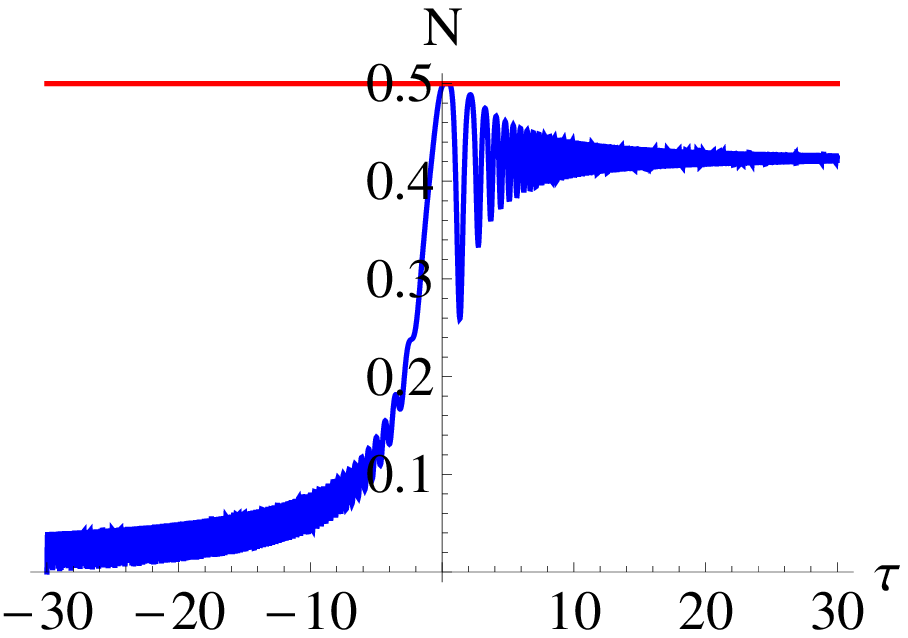}\label{fig:N4beta05}}
\qquad
\subfloat[][]{\includegraphics[width=0.22\textwidth]{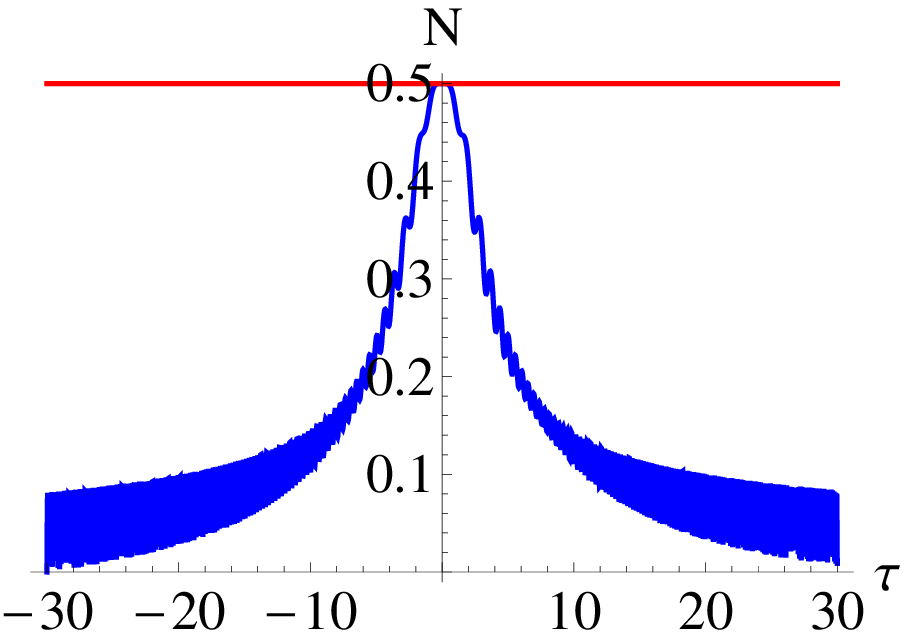}\label{fig:N4beta2}}
\captionsetup{justification=raggedright,format=plain,skip=4pt}%
\caption{(Color online) a) $\beta$-dependence of the asymptotic Negativity of the two qutrits [Eqs. \eqref{Neg 4x4} and \eqref{x}] for the initial condition $\ket{-10}$. Time behaviour of the Negativity against the dimensionless parameter $\tau=\sqrt{\alpha/\hbar}~t$ during a LMSZ process when the two-qutrit system starts from the state $\ket{-10}$ for $2\beta_-=\beta_+$ and b) $\beta_+=0.11$, c) $\beta_+=1/2$ and d) $\beta_+=2$. The upper straight curve represents $\mathcal{N}=0.5$.}
\end{center}
\end{figure}

If we consider as initial condition the two-qutrit state $\ket{-10}$, through the exact form of the time evolution operator in Eq. \eqref{U+}, it is easy to verify that
\begin{equation}\label{xt}
\begin{aligned}
x(t)&=|w_1(t)|^2+|w_4(t)|^2=|a_1|^2|a_2|^2+|b_1|^2|b_2|^2
\end{aligned}
\end{equation}
At infinite time so we have
\begin{equation}\label{x}
x(\infty)=P_1P_2+(1-P_1)(1-P_2),
\end{equation} 
where the expressions of $P_1$ and $P_2$ are reported in Eq. \eqref{P1} and \eqref{P2}, respectively.
If we put the expression in Eq. \eqref{x} into Eq. \eqref{Neg 4x4}, we get the asymptotic expression of the Negativity.
In Fig. \ref{fig:PbetaCurve4} such an expression of the negativity is reported against the LMSZ parameter $\beta=\beta_+$, for $\beta_-/\beta_+=1/2$.
We see that two maxima are present and they correspond to the values $\log(2)/2\pi\approx 0.11$ and $\log(2)/\pi\approx 0.22$.
It means that, by appropriately setting the parameter $\beta$, when the two-qutrit system start from the state $\ket{-10}$, through the LMSZ process we may generate asymptotically an entangled state of the two spin-qutrits with the maximum level of entanglement possible in such a subspace.
This fact is confirmed by Fig. \ref{fig:N4beta011} where the time behaviour of the Negativity is reported against the dimensionless parameter $\tau=\sqrt{\alpha/\hbar}~t$ for $\beta=0.11$.
In this case, we used the expression of $x(t)$ in Eq. \eqref{xt} with the exact solution of the LMSZ dynamical problem which read \cite{Vit-Garr}
\begin{equation}\label{Exact a b}
\begin{aligned}
a_{1/2}=&{\Gamma_f(1-i\beta_\pm) \over \sqrt{2\pi}} \\
\times&[D_{i\beta_\pm}(\sqrt{2}e^{-i\pi/4}\tau)*D_{-1+i\beta_\pm}(\sqrt{2}e^{i3\pi/4}\tau_i) \\
&+D_{i\beta_\pm}(\sqrt{2}e^{i3\pi/4}\tau)*D_{-1+i\beta_\pm}(\sqrt{2}e^{-i\pi/4}\tau_i)],
\\\\
b_{1/2}=&{\Gamma_f(1-i\beta_\pm) \over \sqrt{2\pi\beta}} e^{i\pi/4} \\
\times&[-D_{i\beta_\pm}(\sqrt{2}e^{-i\pi/4}\tau)*D_{-1+i\beta_\pm}(\sqrt{2}e^{i3\pi/4}\tau_i) \\
&+D_{i\beta_\pm}(\sqrt{2}e^{i3\pi/4}\tau)*D_{-1+i\beta_\pm}(\sqrt{2}e^{-i\pi/4}\tau_i)].
\end{aligned}
\end{equation}
$\Gamma_f$ is the gamma function, $D_\nu(z)$ are the parabolic cylinder functions \cite{Abramowitz} and $\tau_i$ identifies the initial time instant.
We emphasize that the parameter $\beta$, besides the asymptotic value, deeply influences the trend in time of the Negativity curve, as it can be appreciated by Figs. \ref{fig:N4beta05} and \ref{fig:N4beta2}, related to $\beta=0.5$ and $\beta=2$, respectively.

It is interesting to point out that the initial state $(\ket{-10}+\ket{0-1})/\sqrt{2}$ under the condition $\omega_1(t)=\omega_2(t)$, we have taken into account in Sec. \ref{Local Fields}, exhibits a constant maximum level of entanglement (1/2) during the evolution.
Such a peculiar feature is independent of the specific time-dependence of the field and it may be understood at the light of the analysis reported in \cite{GMIV}.
There the authors analyse the same two-qutrit model but with time-independent fields.
They have brought to light the existence of eight states with such a feature which is related to the symmetry property of the Hamiltonian.
Being such property unaffected by a general time-dependence of the applied field as we showed before, we find of course the same feature here too.

We stress that it is not possible to get physical information about the entanglement get established between the two qutrits by studying correlations emerging between the two fictitious qubits.
Indeed, by the mapping in Eq. \eqref{Mapping}, it is easy to see that entangled states of the two qutrits, such as $(\ket{10}+\ket{01})/\sqrt{2}$, correspond to separable states of the two qubits, $(\ket{++}+\ket{+-})/\sqrt{2}$, and, \textit{vice versa}, separable states of the qutrits $(\ket{10}+\ket{-10})/\sqrt{2}$ correspond to entangled states of the qubit system, $(\ket{++}+\ket{--})/\sqrt{2}$.
Such a feature stems from the non-locality of the mapping established between the two systems.
This observation implies that, within the four-dimensional subspace, we cannot use the Concurrence, but we are obliged to consider another Entanglement measure.
This is why we use Negativity to quantify the Entanglement get established between the two qutrits.

\subsection{Three-Dimensional Sub-Dynamics}

\begin{figure}[htp]
\begin{center}
\subfloat[][]{\includegraphics[width=0.22\textwidth]{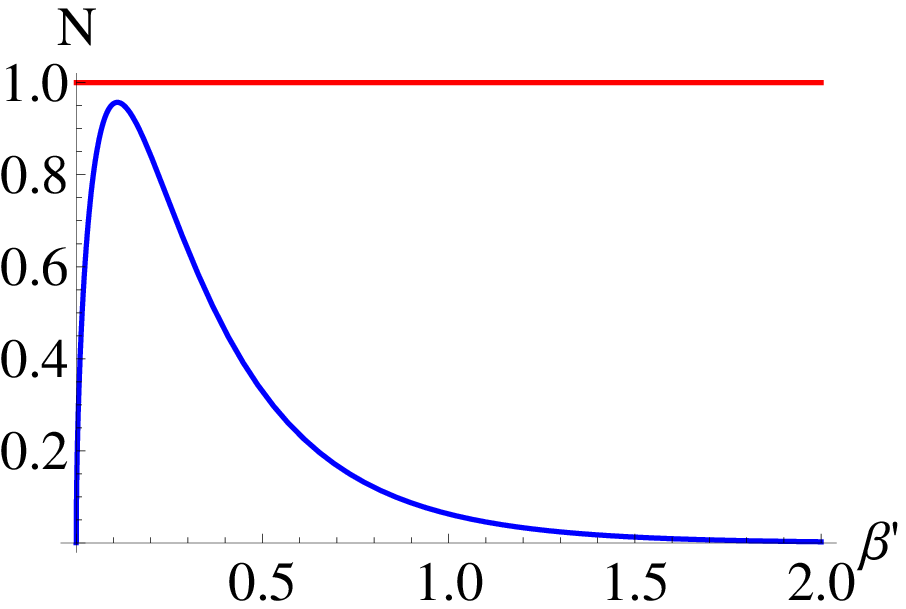}\label{fig:NBetaCurve3}}
\qquad
\subfloat[][]{\includegraphics[width=0.22\textwidth]{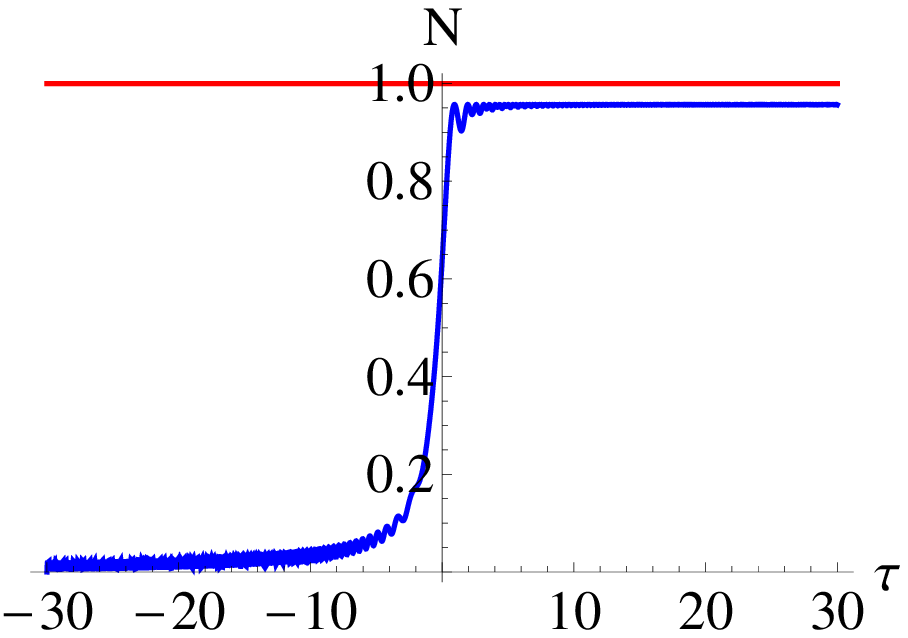}\label{fig:N3beta011}}
\qquad
\subfloat[][]{\includegraphics[width=0.22\textwidth]{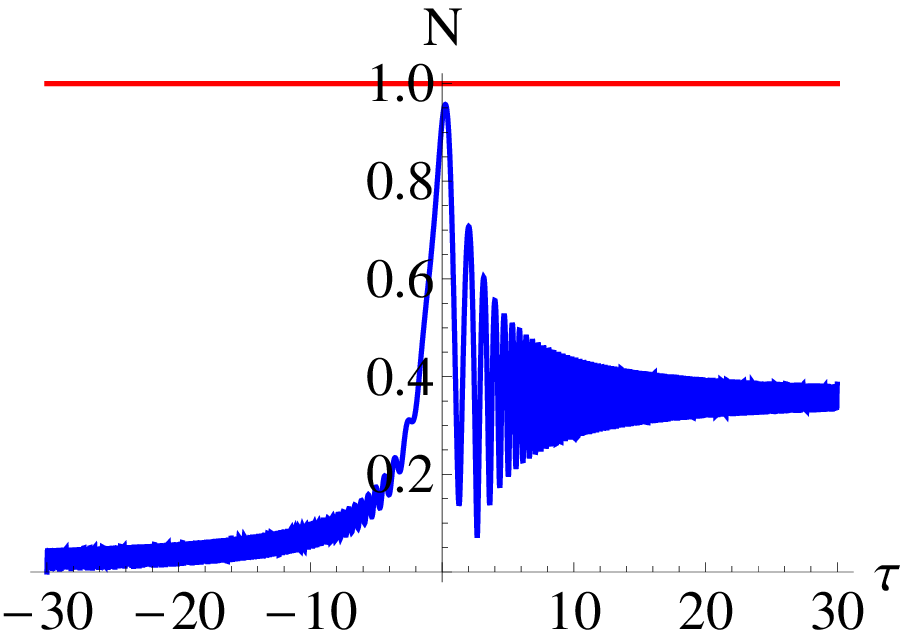}\label{fig:N3beta05}}
\qquad
\subfloat[][]{\includegraphics[width=0.22\textwidth]{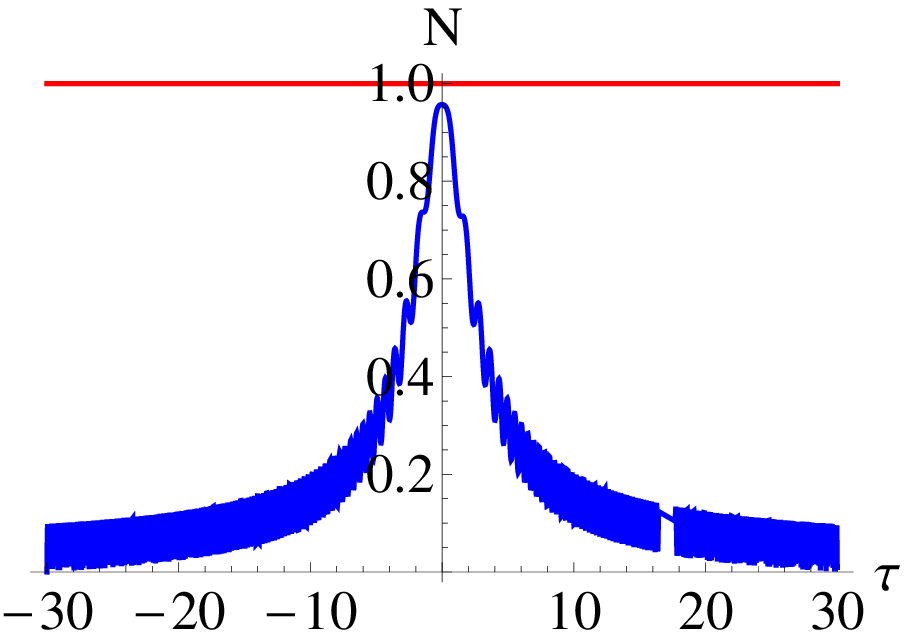}\label{fig:N3beta2}}
\captionsetup{justification=raggedright,format=plain,skip=4pt}%
\caption{(Color online) a) $\beta'$-dependence of the asymptotic Negativity of the two qutrits [Eqs. \eqref{Neg time 3 lev}] for the initial condition $\ket{-11}$. Time behaviour of the Negativity against the dimensionless parameter $\tau=\sqrt{\alpha/\hbar}~t$ during a LMSZ process when the two-qutrit system starts from the state $\ket{-11}$ for b) $\beta'=0.11$, c) $\beta'=1/2$ and d) $\beta'=2$. The upper straight curve represents $\mathcal{N}=0.5$.}
\end{center}
\end{figure}

In the three-dimensional subspace the Negativity for the general state in Eq. \eqref{Gen State 3dim Space} reads
\begin{equation}
\mathcal{N}=|c_1||c_2|+|c_2||c_3|+|c_1||c_3|.
\end{equation}
Its time evolution related to the initial condition $\ket{-11}$ results
\begin{equation}\label{Neg time 3 lev}
\mathcal{N}(t)=|a_3||b_3|[\sqrt{2}+|a_3||b_3|],
\end{equation}
and then asymptotically we get
\begin{equation}
\mathcal{N}(\infty)=P_3(1-P_3)+\sqrt{2P_3(1-P_3)},
\end{equation}
where $P_3$ is defined after Eqs. \eqref{LMSZ Tr Pr 3-lev}.
This quantity reaches its maximum value for $P_3=1/2$ and then for $\beta'=\log(2)/2\pi \approx 0.11$ (see Fig. \ref{fig:NBetaCurve3}).
This means that, for such a value of the parameter $\beta'$, the LMSZ process generates asymptotically an entangled state of the two qutrits with the maximum available value of Negativity for the initial condition under scrutiny, as confirmed by Fig. \ref{fig:N3beta011}.
We got the latter figure by putting in Eq. \eqref{Neg time 3 lev} the expressions of $a_+$ and $b_+$ (or, equivalently, $a_-$ and $b_-$) in Eqs. \eqref{Exact a b}, replacing $\beta_+$ ($\beta_-$) with $\beta'$.
In the same way we have analysed the time behaviour of the Negativity for the same initial condition for other two values of the parameter $\beta'$, namely $\beta'=1/2$ (Fig. \ref{fig:N3beta05}) and $\beta'=2$ (Fig. \ref{fig:N3beta2}).
Also this time we find that the LMSZ parameter deeply influences not only the asymptotic value but also the trend in time of the Negativity.

\section{Conclusive Remarks} \label{Conclusions}

This paper investigates the quantum dynamics of  two interacting qutrits subjected to local time-dependent fields.
We have taken into account the anisotropic as well as isotropic Heisenberg interaction.
The field applied on just one of the two qutrits or on both the two spin-1's has been considered linearly varying on time (LMSZ ramp) along the quantization $z$-axis.
Atomic species with three metastable levels may be used in a linear ion crystal to realize the interacting spin-1 model under scrutiny through the application of laser fields \cite{Cohen2014,Senko2015}.
Moreover, a broad range of physical situations may be covered by such a model: two spin-1's in a double well optical lattice \cite{Yip}, interacting spin-1 nanomagnets \cite{HeXuLiang} and effective interaction between two separated nitrogen-vacancy centres in diamond \cite{Bermudez2011}.

The dynamical problem has been solved thank to the reduction to two easier problems: one of two non-interacting fictitious spin-1/2's and the other of a fictitious three-level system.
Such a reduction relies on the symmetry-based analysis of the Hamiltonian model reported in Ref. \cite{GMIV} which is unaffected by the time-dependences of the applied fields and, more generally, by the time-dependences of all Hamiltonian parameters.
This means that the same analysis may be developed considering other possible time-dependences of the field leading to exactly solvable problems \cite{KunaNaudts,Bagrov,Das Sarma,Mess-Nak,GMN,MGMN,GdCNM}.

The main result of the paper is the physical effect we called coupling-driven LMSZ transition.
It consists in the fact that, though a transverse constant field is absent, LMSZ transitions between two-qutrit states are still possible thanks to the presence of the coupling between the two spin-1's.
Indeed, the fictitious dynamics of the two decoupled qubits and the one of a fictitious spin-1 are characterized by a LMSZ longitudinal field and a fictitious constant transverse field stemming from the coupling existing between the spin-qutrits.
This fact implies that, avoided crossings in the two qutrit system are possible thanks to the presence of such an interaction.
A remarkable consequence of this circumstance consists in the fact that an appropriate ratio between the applied fields and the coupling parameters may result favourable for performing adiabatic dynamics with consequent full LMSZ transitions of the two spin-1 system.
The knowledge of such a physical effect makes it possible to have control on the dynamics of the system under scrutiny as well as to get information about the interaction characterizing the same system.
We have brought to light, moreover, how the LMSZ transition probabilities change according to the (an)isotropy of the coupling terms.

We have showed that the physical relevance of the coupling-driven LMSZ transitions is twofold.
Firstly, by the knowledge of the transition probabilities we may estimate the coupling parameters of the two-qutrit model.
Secondly, basing on such an estimation, we illustrated that an appropriate and specific choice of the slope of the LMSZ ramp can generate asymptotically entangled states of the two qutrits.
We have analysed the level of entanglement by studying both the asymptotic Negativity against the LMSZ parameters and its time evolution.
In the latter case, we have used the exact solutions of the LMSZ dynamical problem \cite{Vit-Garr} and we have investigated the effects of the coupling determining the LMSZ parameter.
We reported how such a parameter, depending on the ratio of the squared coupling and the slope of the ramp, determines not only the asymptotic value, but also the trend of the Negativity.

Finally, we have discussed also how the LMSZ transition probabilities are modified by the presence of a noisy field component stemming from the interaction of the the two-qutrit system with a surrounding environment.
Such an analysis is based on the fact that the dynamical reduction is unaffected by the presence of the noise and so, also in this case, we may reduce the two-spin-1 problem to easier problems whose solutions are known in literature.
Following the same philosophy, we have exposed the possibility of treating exactly the problem also by introducing the environment effects with non-Hermitian terms in the Hamiltonian model.

We note that the parameters of the applied magnetic field, including the 
magnetic field gradient, can be controlled in very wide ranges. For 
example, the magnetic field gradient can reach values as large as 
150-200 T/m in a microfabricated ion trap \cite{Weidt},
which is far beyond what is needed here. The most important 
parameter for the feasibility of our scheme is the spin-spin coupling 
constant $\gamma$. In nuclear magnetic resonance, its values typically vary 
from 10 Hz to 300 Hz depending on the molecule \cite{Chuang},
which implies that entanglement can be created on the 
millisecond  scale. A very interesting physical platform, which allows 
the tuning of the spin-spin coupling in a broad range, is provided by 
microwave-driven trapped ions in the presence of a static magnetic-field 
gradient \cite{Weidt,Johanning}. The effective spin-spin coupling is 
proportional to the magnetic-field gradient and can reach the kHz range. 
A third example is provided by Rydberg atoms and ions where, due to the 
huge electric-dipole moments of the Rydberg states, the effective 
spin-spin coupling can reach a few MHz \cite{Urban,Gaetan,Muller}. This 
implies entanglement creation on the sub-microsecond scale.

We do emphasize that the results achieved in this paper are not simple generalizations of the ones reported in \cite{GVM} where the Hamiltonian model \eqref{Hamiltonian} has been investigated for two interacting qubits.
Consider indeed that the symmetry-based existence of two dynamically invariant subspaces regardless of the values of the two spins, does not represent the successful key of our approach in its own.
It is in fact possible to persuade oneself that the effective quantum dynamics of two interacting qudits, restricted to one of the two dynamically invariant subspaces, turns out to be in general a challenging problem whose difficulty grows with increasing the values of the qudits.
This paper shows that, in the quantum dynamics of two qutrits restricted to the two invariant subspaces, these difficulties can successfully overcome by establishing a direct link with su(2) problems. 
Summing up, the route followed in this paper has the merit of explicitly showing that the resolution of the related dynamical problems cannot be derived simply generalizing technical aspects characterizing the analogous dynamical problem of two qubits \cite{GVM}.

It is worth noticing that the ideas and tools which our approach hinges upon may be useful to investigate other even more complex physical scenarios, as for example done in Ref.s \cite{GBNM,GLSM}.
We feel that our results might stimulate possible experimental investigations, for example within STM scenarios \cite{Sivkov}.

Finally, the main perspective of this work is to take into account also quantum degrees of freedom of the bath.
In this case, the basic and fundamental symmetry-based dynamical reduction might be joined with recent approaches \cite{SHGM} to reach a deeper understanding of the dynamics of two-qutrit systems in more realistic experimental situations.

\section{Acknowledgements}
NVV acknowledges support  from the EU Horison-2020 project 820314 (MicroQC).
RG acknowledges economical supports by research funds difc 3100050001d08+, University of Palermo.

\end{document}